\documentclass[10pt,twocolumn,twoside]{IEEEtran}
\usepackage{graphicx}
\usepackage{amsmath}
\usepackage{array}
\newcommand{\PreserveBackslash}[1]{\let\temp=\\#1\let\\=\temp}
\newcolumntype{C}[1]{>{\PreserveBackslash\centering}p{#1}}
\newcolumntype{R}[1]{>{\PreserveBackslash\raggedleft}p{#1}}
\newcolumntype{L}[1]{>{\PreserveBackslash\raggedright}p{#1}}
\usepackage[usenames]{color}
\usepackage{colortbl,booktabs}
\usepackage{tabularx}
\usepackage{multicol}
\usepackage{booktabs}
\usepackage[Symbol]{upgreek}
\usepackage{subfigure}
\usepackage{bm}
\usepackage{cite}
\usepackage{balance}
\usepackage[boxed]{algorithm2e}
\usepackage{url}

\usepackage{threeparttable}
\usepackage{amsmath}
\usepackage{dcolumn}
\usepackage{multirow}
\usepackage{booktabs}
\usepackage{amsfonts}
\newcolumntype{d}[1]{D{.}{.}{#1}}% or D{.}{,}{#1} or D{.}{\cdot}{#1}

\begin{document}

\bibliographystyle{IEEEtran} % use IEEEtran.bst style
\title{Low RF-Complexity Technologies to Enable Millimeter-Wave MIMO with Large Antenna Array for 5G Wireless Communications}

\author{Xinyu Gao,~\IEEEmembership{Student Member,~IEEE}, Linglong Dai,~\IEEEmembership{Senior Member,~IEEE}, and Akbar M. Sayeed,~\IEEEmembership{Fellow,~IEEE}

\thanks{This work has been posted on arXiv (arXiv:1607.04559) on 15 July, 2016 and revised on 16 Sep, 2017. It has also been posted on the personal website at: http://oa.ee.tsinghua.edu.cn/dailinglong/publications/publications.html.}
\thanks{X. Gao and L. Dai (corresponding author) are with the Tsinghua National Laboratory
for Information Science and Technology (TNList), Department of Electronic Engineering, Beijing 100084, China (e-mails: xy-gao14@mails.tsinghua.edu.cn, daill@tsinghua.edu.cn).}
\thanks{A. Sayeed is with the Department of Electrical and Computer Engineering, University of Wisconsin, Madison, WI 53706, USA (email: akbar@engr.wisc.edu).}
\thanks{This work was supported by the National Natural Science Foundation of China for Outstanding Young Scholars (Grant No. 61722109), the National Natural Science Foundation of China (Grant No. 61571270), the Royal Academy of Engineering through the UK-China Industry Academia Partnership Programme Scheme (Grant No. UK-CIAPP${\backslash}$49), and the US National Science Foundation (Grant Nos. 1548996 and 1444962).}}

\maketitle
\begin{abstract}
Millimeter-wave (mmWave) MIMO with large antenna array has attracted considerable interests from academic and industry communities, as it can provide larger bandwidth and higher spectrum efficiency. However, with hundreds of antennas, the number of radio frequency (RF) chains required by mmWave MIMO is also huge, leading to unaffordable hardware cost and power consumption in practice. In this paper, we investigate low RF-complexity technologies to solve this bottleneck. We first review the evolution of low RF-complexity technologies from microwave frequencies to mmWave frequencies. Then, we discuss two promising low RF-complexity technologies for mmWave MIMO systems in detail, i.e., phased array based hybrid precoding (PAHP) and lens array based hybrid precoding (LAHP), including their principles, advantages, challenges, and recent results. We compare the performance of these two technologies to draw some insights about how they can be deployed in practice. Finally, we conclude this paper and point out some future research directions in this area.
\end{abstract}

\section{Introduction}\label{S1}

\IEEEPARstart Millimeter-wave (mmWave) (30-300 GHz) multiple-input multiple-output (MIMO) with large antenna array has been considered as a promising solution to meet the one thousand times increase in data traffic predicted for further 5G wireless communications~\cite{mumtaz2016mmwave}. On one hand, mmWave can provide nearly 2 GHz bandwidth~\cite{rappaport2013millimeter}, which is much larger than the 20 MHz bandwidth in current 4G wireless communications without carrier aggregation. On the other hand, the short wavelengths associated with mmWave frequencies enable a large antenna array to be packed in a small physical size, which means that MIMO with a large antenna array is possible at mmWave frequencies to effectively compensate the high path loss induced by high frequencies and considerably improve the spectrum efficiency~\cite{heath2015overview}.

However, realizing mmWave MIMO in practice is not a trivial task. One challenging problem is that each antenna in MIMO systems usually requires one dedicated radio-frequency (RF) chain, including digital-to-analog converters (DACs), mixers, and so on~\cite{heath2015overview}. This will result in unaffordable hardware cost and power consumption in mmWave MIMO systems, as the number of antennas is huge (e.g., 256 compared with 8) and the power consumption of RF chain is high (e.g., 250 mW at mmWave frequencies compared with 30 mW at microwave frequencies)~\cite{heath2015overview}. Therefore, the large number of RF chains with prohibitively high power consumption is a bottleneck for mmWave MIMO with large antenna array in practice~\cite{mumtaz2016mmwave,heath2015overview}.

In this paper, we investigate low RF-complexity technologies for mmWave MIMO systems. Our contributions can be summarized as follows: i) We provide a review for the evolution of low RF-complexity technologies from microwave frequencies to mmWave frequencies, and highlight two promising technologies proposed recently, i.e., phased array based hybrid precoding (PAHP)~\cite{el2013spatially} and lens array based hybrid precoding (LAHP)~\cite{brady2013beamspace}. We give detailed overview for these two technologies including their principles, advantages, challenges, and recent results; ii) We propose a novel adaptive selecting network for LAHP with low hardware cost and power consumption. For data transmission, it can select beams like the traditional one, while for channel estimation, it can formulate the beamspace channel estimation as a sparse signal recovery problem and estimate the beamspace channel with considerably reduced pilot overhead; iii) We provide the sum-rate and power efficiency comparisons between PAHP and LAHP in a practical outdoor mmWave MIMO system, where the channel estimation error and inter-cell interference are also included. Then, we draw some insights about how these two technologies can be deployed in practice.

\section{Traditional Low RF-Complexity Technologies}\label{S2}
We first review two typical low RF-complexity technologies, i.e., antenna selection and analog beamforming. Antenna selection may be considered as the most classical low RF-complexity technology for microwave MIMO systems. By contrast, analog beamforming is the most widely used low RF-complexity technology for indoor mmWave communications. These two technologies can be regarded as the basics of PAHP and LAHP.

\begin{figure}[tp]
\begin{center}
\vspace*{-1mm}\includegraphics[width=0.95\linewidth]{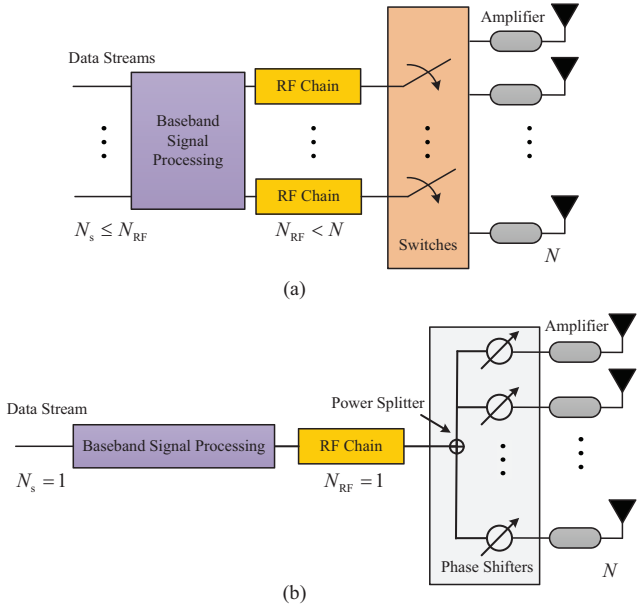}
\end{center}
\vspace*{-4mm}\caption{Architectures of traditional low RF-complexity technologies: (a) antenna selection; (b) analog beamforming.} \label{FIG1}
\end{figure}

\subsection{Antenna selection}\label{S2.1}
As shown in Fig. 1 (a), the key feature of antenna selection~\cite{molisch2004mimo} is that there is one selecting network between ${{N_{{\rm{RF}}}}}$  RF chains and ${N}$  antennas. Based on the channel state information (CSI), the target of antenna selection is to select  ${{N_{{\rm{RF}}}}}$ best antennas out of total ${N}$  antennas for data transmission to maximize the achievable sum-rate~\cite{molisch2004mimo}.

An exciting result of antenna selection is that when the number of RF chains ${{N_{{\rm{RF}}}}}$   is larger than the number of transmitted data streams ${{N_{\rm{s}}}}$, the performance loss induced by antenna selection is negligible under independent identically distributed (IID) Rayleigh fading channels~\cite{molisch2004mimo}. However, when channels are highly correlated, the achievable sum-rate of antenna selection will decrease drastically~\cite{molisch2004mimo}, as antenna selection incurs more channel information loss in this case.

%\subsection{Spatial modulation}\label{S2.2}
%Spatial modulation can be regarded as an evolution of antenna selection, where the key improvement is that the spatial index of each antenna can be exploited as an extra source for data transmission~\cite{di2014spatial}. As shown in Fig. 1 (b), the transmitted data streams are first divided into two blocks by the spatial modulation mapper. The bits in the first block are used to determine which antennas are active for data transmission, while the bits in the second block are passed through the baseband signal processing module with conventional constellation modulation, precoding, and so on. Since the signals transmitted by different antennas are expected to experience different propagation channels, at the receiver, we can detect not only the transmitted signals, but also which antennas are active if CSI is known. After that, the transmitted bits in both two blocks above can be recovered simultaneously.
%
%The main advantage of spatial modulation is that it can provide extra increase in sum-rate without expanding the bandwidth~\cite{di2014spatial}. However, as only ${{N_{{\rm{RF}}}}}$  out of  ${N}$ antennas (${{N_{{\rm{RF}}}} \ll N}$) are active at the same time, the antenna gain achieved by spatial modulation is limited. Moreover, if the channels of active antennas are not sufficiently different, we cannot accurately distinguish which antennas are used at the receiver, and spatial modulation will suffer from serious performance loss~\cite{di2014spatial}.

\subsection{Analog beamforming}\label{S2.3}
As shown in Fig. 1 (b), the key idea of analog beamforming~\cite{hur2013millimeter} is to use only one RF chain to transmit single data stream, and employ the phase shifter network to control the phases of original signals to maximize the array gain and effective signal-to-noise ratio (SNR). Building the beamforming vectors (include the relative amplitudes and phases applied to the different antenna elements to shape the signal strength at a specific direction in the far field) requires beam training, which involves an iterative and joint design between the transmitter and receiver. For example, in IEEE 802.11ad, a multi-resolution beamforming codebook (consists of several pre-defined beamforming vectors) is adopted to progressively refine the selected beamforming vectors~\cite{hur2013millimeter}.

The advantage of analog beamforming is that it only requires one RF chain, leading to quite low hardware cost and power consumption~\cite{hur2013millimeter}. However, the analog beamforming can only adjust the phases of the signals, which means that all the elements of beamforming vector have the same amplitude. Such design constraint will incur some performance loss~\cite{heath2015overview}. More importantly, analog beamforming can only support single-stream transmission, which cannot be used in multi-stream or multi-user scenarios~\cite{heath2015overview}.

\subsection{Can traditional low RF-complexity technologies be used in mmWave MIMO systems?}\label{S3.3}
Note that the outdoor mmWave MIMO channel is significantly different from the one at microwave frequencies. One of the most important differences is that the scattering of outdoor mmWave communications is usually limited. This is because that the wavelengths at mmWave frequencies are quite small compared to the obstacle size, leading to poor diffraction~\cite{heath2015overview}. Moreover, though scattering occurs, it incurs significant 5-20 dB attenuation at mmWave frequencies~\cite{rappaport2013millimeter,brady2013beamspace}. A study in New York has shown that the average number of paths in 28 GHz outdoor mmWave communications is only 2.4~\cite{rappaport2013millimeter}. This means that the outdoor mmWave MIMO channel is usually low-rank with high correlation in the spatial domain and sparse in the angular domain~\cite{brady2013beamspace}.

Based on these facts, we know that antenna selection is not appropriate for mmWave MIMO systems~\cite{rappaport2013millimeter}, since it suffers from serious performance loss with highly correlated channels. Moreover, although analog beamforming is developed for mmWave communications, it can only support single-stream transmission without multiplexing gains. This means that analog beamforming cannot fully exploit the potential of mmWave MIMO in spectrum efficiency~\cite{heath2015overview,brady2013beamspace}. Next, we will investigate two promising low RF-complexity technologies proposed recently for mmWave MIMO systems, i.e., PAHP and LAHP.

\section{Phased Array Based Hybrid Precoding}\label{S4}

%\begin{figure}[tp]
%\begin{center}
%\vspace*{-1mm}\hspace*{-1mm}\includegraphics[width=0.95\linewidth]{hybrid1}
%\end{center}
%\vspace*{-4mm}\caption{Architecture comparison: (a) fully digital precoding; (b) full-PAHP; (c) sub-PAHP.} \label{FIG1}
%\end{figure}

\subsection{Principle}\label{S4.1}
Precoding is used to adjust the weights of transmitted signals to maximize the achievable sum-rate~\cite{heath2015overview}. As shown in Fig. 2 (a), the conventional fully digital precoding can arbitrarily adjust the amplitudes and phases of the original signals. It can achieve multiplexing gains, and enjoys higher design freedom than analog beamforming. However, it requires one dedicated RF chain for each antenna, which brings unaffordable hardware cost and power consumption when the number of antennas is large. Hybrid precoding can be considered as a promising compromise between the optimal fully digital precoding and the low-cost analog beamforming~\cite{el2013spatially,brady2013beamspace}. Its key idea is to divide the large-size digital precoder into a large-size analog beamformer (realized by the analog circuit) and a small-size digital precoder (requiring a small number of RF chains).

PAHP is one of the realization of hybrid precoding, where the analog beamformer is realized by phase shifters. Assume there are ${{N_{\rm{s}}}}$ single-antenna users to be served. As shown in Fig. 2, the received signal vector ${{\bf{y}}}$ for ${{N_{\rm{s}}}}$ users in the downlink can be presented as
\begin{equation}\label{eq1}
{\bf{y}} = {\bf{HADs}} + {\bf{n}},
\end{equation}
where ${{\bf{H}}}$ of size ${{N_{\rm{s}}} \times N}$, ${{\bf{s}}}$ of size ${{N_{\rm{s}}} \times 1}$, and ${{\bf{n}}}$ of size ${{N_{\rm{s}}} \times 1}$ denote the mmWave MIMO channel matrix, transmitted signal vector, and noise vector, respectively, ${{\bf{A}}}$ of size ${N \times {N_{{\rm{RF}}}}}$ is the analog beamformer, and ${{\bf{D}}}$ of size ${{N_{{\rm{RF}}}} \times {N_{\rm{s}}}}$ is the digital precoder. Note that PAHP has two architectures, i.e., full-PAHP~\cite{el2013spatially} and sub-PAHP~\cite{gao15energy}, as illustrated in Fig. 2 (b) and Fig. 2 (c), respectively. In full-PAHP, each RF chain is connected to all ${N}$ antennas via phase shifters, and the analog beamformer ${{\bf{A}}}$ is a full matrix. It can achieve satisfying performance, but usually requires a large number of ${N{N_{{\rm{RF}}}}}$  phase shifters (e.g., ${N{N_{{\rm{RF}}}} = 256 \times 16 = 4096}$), together with the complicated power splitters/cominbers and signal/control lines. By contrast, in sub-PAHP, each RF chain is only connected to a subset of antennas, leading ${{\bf{A}}}$  to be a block diagonal matrix. Obviously, sub-PAHP can reduce the number of phase shifters from ${N{N_{{\rm{RF}}}}}$ to ${N}$ and avoid using power combiners~\cite{gao15energy}. Therefore, although sub-PAHP suffers from a loss in array gains by a factor of ${1/{N_{{\rm{RF}}}}}$, it may be preferred in practice~\cite{gao15energy}.

\vspace*{-1mm}
\subsection{Advantages}\label{S4.2}
PAHP can achieve a better tradeoff between the hardware cost/power consumption and the sum-rate performance. It can significantly reduce the number of required RF chains from ${N}$ (e.g., ${N = 256}$) to ${{N_{\rm{s}}}}$ (e.g., ${{N_{{\rm{RF}}}} = {N_{\rm{s}}} = 16}$), leading to lower  power consumption. Besides, as explained above, the outdoor mmWave MIMO channel matrix is usually low-rank~\cite{rappaport2013millimeter}. This indicates that the maximum number of data streams that can be simultaneously transmitted by such channel is limited. Therefore, as long as the number of RF chains is larger than the rank of channel matrix, the small-size digital precoder is still able to fully achieve the multiplexing gains and obtain the near-optimal performance compared to the fully digital precoding~\cite{el2013spatially,gao15energy}.

\vspace*{-1mm}
\subsection{Challenges and recent results}\label{S4.3}
\emph{\textbf{Optimal design of hybrid precoder}}: Maximizing the achievable sum-rate by designing the hybrid precoder ${{\bf{P}} = {\bf{AD}}}$ is the main target of PAHP. However, this optimization problem imposes new challenges, since there are several non-convex hardware constraints on the analog beamformer ${{\bf{A}}}$. For example, all the nonzero elements of ${{\bf{A}}}$  should share the same amplitude due to the constant modulus constraint on phase shifter. To this end, one feasible way is to approximate the original optimization problem as a convex one to obtain a near-optimal hybrid precoder with low complexity.

Following this idea, some advanced schemes have been proposed recently. In~\cite{el2013spatially}, a spatially sparse scheme is proposed for single-user full-PAHP. It approximates the sum-rate optimization problem as the one minimizing the distance between the optimal fully digital precoder and the hybrid precoder. Then, a variant of the orthogonal matching pursuit (OMP) algorithm~\cite{bajwa2010compressed} is developed to obtain the near-optimal hybrid precoder. In~\cite{alkhateeb2015limited}, full-PAHP is extended to multi-user scenario, where a two-stage multi-user scheme is proposed. In the first stage, the optimal analog beamformer is searched from a pre-defined codebook to maximize the desired signal power of each user. In the second stage, the classical zero forcing (ZF) precoder is used to cancel multi-user interference. In~\cite{gao15energy}, a successive interference cancelation (SIC)-based scheme is proposed for sub-PAHP. It first decomposes the sum-rate optimization problem into a series of simple and convex sub-problems, each of which only considers one sub-phased array. Then, inspired by the classical SIC multi-user signal detector, the near-optimal hybrid precoder for each sub-phased array is obtained in an one-by-one fashion.

\vspace*{+2mm}
\emph{\textbf{Channel estimation}}: The maximum gain of PAHP can be only achieved with perfect CSI, which is difficult to obtain in mmWave MIMO systems. Firstly, due to the lack of array gains before the establishment of the transmission link, the SNR for channel estimation in PAHP is quite low~\cite{heath2015overview}. Secondly, the number of RF chains in PAHP is usually much smaller than the number of antennas. Therefore, we cannot directly observe the channel matrix like that in fully digital precoding.

Two typical solutions have been proposed to solve this problem. The first divides the channel estimation problem into two steps. In the first step, the BS and users will perform beam training like analog beamforming~\cite{hur2013millimeter} to determine ${{\bf{A}}}$. In the second step, the effective channel matrix ${{\bf{HA}}}$ with smaller size ${{N_{\rm{s}}} \times {N_{{\rm{RF}}}}}$ (${{N_{{\rm{RF}}}} \ll N}$) is estimated by classical schemes, such as least squares (LS).  The second solution is to exploit the low-rank characteristic of mmWave MIMO channel. Instead of estimating the effective channel matrix ${{\bf{HA}}}$, it can directly obtain the complete channel matrix ${{\bf{H}}}$ with low pilot overhead. For example, an adaptive compressive sensing (CS)~\cite{bajwa2010compressed} based channel estimation scheme is proposed in~\cite{heath2015overview}. It divides the total channel estimation problem into several sub-problems, each of which only considers one channel path. For each channel path, it first starts with coarse direction grids, and determines the direction of this path belonging to which grid by employing OMP algorithm. Then, the narrowed direction grids are used,  and the direction of this path is further refined. Note that the first solution usually involves low complexity and is easy to implement. By contrast, the second solution can significantly reduce the pilot overhead, but usually involves higher complexity.

\section{Lens Array Based Hybrid Precoding}\label{S5}

%\begin{figure}[tp]
%\begin{center}
%\vspace*{-1mm}\includegraphics[width=0.95\linewidth]{beamspace1}
%\end{center}
%\vspace*{-4mm}\caption{Architectures of LAHP with: (a) traditional selecting network; (b) proposed adaptive selecting network.} \label{FIG1}
%\end{figure}

\subsection{Principle}\label{S5.1}
Although full-PAHP can achieve the near-optimal performance with reduced number of RF chains, it usually requires a large number of high-resolution phase shifters, together with the complicated power splitters/combiners. Sub-PAHP can partly solve these problems, but it suffers from some performance loss due to the reduced array gains.

These problems above can be solved by LAHP~\cite{brady2013beamspace}, another realization of hybrid precoding, where the analog beamformer is realized by lens array and selecting network as shown in Fig. 3 (a). By employing a lens array (an electromagnetic lens with directional energy focusing capability and a matching antenna array with elements located in the focal surface of the lens~\cite{brady2013beamspace}), the signals from different directions (beams) can be concentrated on different antennas, and the spatial channel can be transformed to the beamspace channel. Mathematically, the lens array plays the role of a spatial discrete fourier transform (DFT) matrix ${{\bf{U}}}$ of size  ${N \times N}$, whose ${N}$ columns correspond to the orthogonal beamforming vectors of ${N}$ pre-defined directions (beams) that cover the whole angular space. The system model of LAHP can be presented by
\begin{equation}\label{eq2}
{\bf{\tilde y}} = {\bf{HUDs}} + {\bf{n}} = {\bf{\tilde HDs}} + {\bf{n}},
\end{equation}
where ${{\bf{\tilde y}}}$, ${{\bf{s}}}$, and ${{\bf{n}}}$ of size ${{N_{\rm{s}}} \times 1}$ denote the received signal vector in the beamspace, transmitted signal vector, and noise vector, respectively, ${{\bf{D}}}$ of size ${N \times {N_{\rm{s}}}}$ is the digital precoder, and the beamspace channel ${{\bf{\tilde H}}}$ of size ${{N_{\rm{s}}} \times N}$ is defined as ${{\bf{\tilde H}} = {\bf{HU}}}$, whose ${N}$  columns correspond to ${N}$  orthogonal beams.

\subsection{Advantages}\label{S5.2}
LAHP can also achieve the near-optimal performance with low hardware cost/power consumption. This advantage comes from the fact that the beamspace channel ${{\bf{\tilde H}}}$ at mmWave frequencies is sparse due to the limited scattering~\cite{rappaport2013millimeter}. Therefore, similar to antenna selection, we can select only a small number of dominant beams to reduce the MIMO dimension as ${{\bf{\tilde y}} \approx {{\bf{\tilde H}}_{\rm{r}}}{{\bf{D}}_{\rm{r}}}{\bf{s}} + {\bf{n}}}$,  where ${{{\bf{\tilde H}}_{\rm{r}}} = {\bf{\tilde H}}{\left( {:,l} \right)_{l \in {\cal B}}}}$ is the dimension-reduced beamspace channel, ${{\cal B}}$  denotes the set of selected beams, and ${{{\bf{D}}_{\rm{r}}}}$ of size ${\left| {\cal B} \right| \times {N_{\rm{s}}}}$  is the corresponding dimension-reduced digital precoder. As the dimension of ${{{\bf{D}}_{\rm{r}}}}$ is much smaller than that of ${{\bf{D}}}$ in (\ref{eq2}), LAHP can significantly reduce the number of required RF chains without obvious performance loss. Another advantage of LAHP is that the array gains can be always preserved by the low-cost lens array. Therefore, even though the simple selecting network is used, the satisfying performance can be still guaranteed~\cite{brady2013beamspace}.

%it is more suitable for fast time-varying channels. Since lens  array can concentrate the signals from different directions on different antennas, when the mmWave MIMO channel varies, the indices of the energy-focusing antennas will change, but the array gains are still preserved. Therefore, we only need to scan the antennas by the selecting network to select the energy-focusing antennas, and then  ${{{\bf{D}}_{\rm{r}}}}$ can be easily re-computed according to the updated ${{{\bf{H}}_{\rm{r}}}}$ with reduced dimension. By contrast, for PAHP, when the channel varies, the previously designed analog beamformer and digital precoder will suffer from non-negligible performance loss. As a result, we have to re-estimate the whole mmWave MIMO channel and re-design the optimal analog beamformer and digital precoder, leading to high complexity.

\subsection{Challenges and recent results}\label{S5.3}
\emph{\textbf{Optimal design of beam selection}}: The performance of LAHP depends on beam selection, which aims to select ${\left| {\cal B} \right|}$ beams out of the total ${N}$  beams to maximize the achievable sum-rate. The most intuitive beam selection scheme is exhaustive search. It has the optimal performance but prohibitively high complexity, which exponentially increases with the number of selected beams. This means that more efficient beam selection schemes should be designed.

In~\cite{brady2013beamspace}, a magnitude maximization (MM) beam selection scheme is proposed, where several beams with large power are selected for data transmission. MM beam selection scheme is simple, but it only aims to preserve the power as much as possible without considering interference, leading to some performance loss. In~\cite{amadorilow}, the authors propose a more efficient beam selection scheme by using the incremental algorithm developed from antenna selection. It selects ${\left| {\cal B} \right|}$ beams one by one. In each step, the beam with the greatest contribution to the achievable sum-rate is selected. In~\cite{gao16bs}, an interference-aware (IA) beam selection scheme with better performance is proposed. The key idea is to classify all users into two user groups according to the potential interference. For users with small interference, it directly selects the beams with large power, while for users with severe interference, the incremental algorithm is employed to search the optimal beams.

\vspace*{+2mm}
\emph{\textbf{Channel estimation}}: The channel estimation for LAHP is different from that for PAHP since: i) PAHP and LAHP have different hardware architectures; ii) we need to estimate the sparse beamspace channel instead of the conventional low-rank spatial channel~\cite{brady2013beamspace}.

This problem can be solved by following two recently proposed solutions. The first one is to reduce the dimension of beamspace channel estimation problem. For example, in~\cite{Hogan16}, a two-step channel estimation scheme is proposed. In the first step, the BS utilizes the selecting network to scan all the beams, and selects several strong beams. In the second step, LS is used to estimate the dimension-reduced beamspace channel. The second solution is to completely estimate the sparse beamspace channel by utilizing CS algorithms. However, if we use the traditional selecting network as shown in Fig. 3 (a) (each RF chain can only select one beam via one switch) to sample the beamspace channel with reduced number of pilots, the sensing matrix will be rank-deficient and CS algorithms cannot work~\cite{bajwa2010compressed}. To this end, we propose an adaptive selecting network for LAHP as shown in Fig. 3 (b), where each RF chain is connected to all antennas via switches. For data transmission, it can select beams like the traditional one, while for channel estimation, it can perform as a sensing matrix with randomly selected 0/1 elements, which has full rank~\cite{bajwa2010compressed}. Then, we can estimate the sparse beamspace channel via efficient CS algorithms such as OMP, and considerably reduce the pilot overhead\footnote{In practice, the beamspace channel may not be sparse enough due to the power leakage effect, but this can be relieved following the idea in~\cite{taubock2010compressive}.}. Note that the extra power consumption and hardware cost incurred by the adaptive selecting network is limited, since the switches are easy to implement with low power consumption~\cite{mendez2016hybrid}. Finally, similar to PAHP, the first solution  usually involves low complexity while the second one can considerably reduce the pilot overhead.

\section{Performance Comparison}\label{S6}
In this section, we compare the performance of full-PAHP, LAHP with the proposed adaptively selecting network, and fully digital precoding. We first consider an outdoor multi-user mmWave MIMO in single-cell scenario, where the BS employs ${N = 256}$  antennas to simultaneously serve ${{N_{\rm{s}}} = 16}$ single-antenna users. The widely used Saleh-Valenzuela multi-path model is adopted to capture the characteristics of mmWave MIMO channels~\cite{rappaport2013millimeter}. Each user has one LoS path and two NLoS paths. The gain of LoS path is normalized to 1, while the gain of each NLoS path is assumed to follow ${{\cal C}{\cal N}\left( {0,0.1} \right)}$. The directions of all paths of users are assumed to follow the IID uniform distribution within ${\left[ { - \pi/2 ,\pi/2 } \right]}$.

%\begin{figure}[tp]
%\begin{center}
%\vspace*{+1mm}\includegraphics[width=0.9\linewidth]{K_factor1}
%\end{center}
%\vspace*{-4mm}\caption{Comparison of sum-rate against Rician factor ${\gamma}$.} \label{FIG1}
%\end{figure}

Fig. 4 shows the comparison of achievable sum-rate against the SNR for data transmission. For full-PAHP with ${{N_{{\rm{RF}}}} = {N_{\rm{s}}} = 16}$ RF chains, we use the adaptive CS based channel estimation scheme proposed in~\cite{heath2015overview} (the number of initial direction grids is set as 1024) to estimate the spatial channel, and employ the two-stage multi-user scheme proposed in~\cite{alkhateeb2015limited} (4-bit phase shifters are used) to transmit data. For LAHP also with ${{N_{{\rm{RF}}}} = {N_{\rm{s}}} = 16}$ RF chains, we utilize the proposed adaptive selecting network to estimate the beamspace channel as illustrated above, and adopt the IA beam selection scheme~\cite{gao16bs} to transmit data. Finally, for fully digital precoding with ${{N_{{\rm{RF}}}} = N = 256}$ RF chains, the classical LS channel estimation scheme and ZF precoding scheme are adopted. For fair comparison, we set SNR as 20 dB and use 96 pilots for all of the three adopted channel estimation schemes. From Fig. 4, we observe that, although the number of RF chains is significantly reduced, both PAHP and LAHP can achieve the sum-rate performance close to that of fully digital precoding. This is due to the fact that PAHP can fully exploit the low-rank characteristic of mmWave channels in the spatial domain, while LAHP can benefit from the sparse characteristic of mmWave channels in the beamspace (angular domain). Since the lens array plays the role of spatial DFT, the beamspace channel and the spatial channel are essentially equivalent but expressed in different forms, just like the same signal can be expressed in the time domain and equivalently in the frequency domain. Moreover, Fig. 4 also shows that with the perfect channel, PAHP outperforms LAHP by about 2 dB, since the phase shifter has higher design freedom than switch. Finally, Fig. 4  shows that LAHP is more robust to the channel estimation error, since only the dimension-reduced beamspace channel is effective for data transmission and a little inaccurate channel may still achieve satisfying  performance.

%Fig. 5 shows the comparison of achievable sum-rate against Rician factor ${\gamma}$ (the ratio
%between the power of LoS path and the average power of NLoS paths~\cite{rappaport2013millimeter}), where the power of LoS path is normalized to 1. From Fig. 5, we observe that with the increased Rician factor ${\gamma}$, the performance of all technologies decreases. This can be explained by the fact that when ${\gamma}$ is high, the power of the total channel will be low. In addition, we also observe that with the increased ${\gamma}$, the performance gap between PAHP and LAHP also decreases. This is due to the fact that when ${\gamma}$ is high, the mmWave MIMO channel is more like a single-path channel, whose power can be more easily reserved after the beam selection in LAHP.

\begin{figure}[tp]
\begin{center}
\vspace*{-1mm}\includegraphics[width=0.95\linewidth]{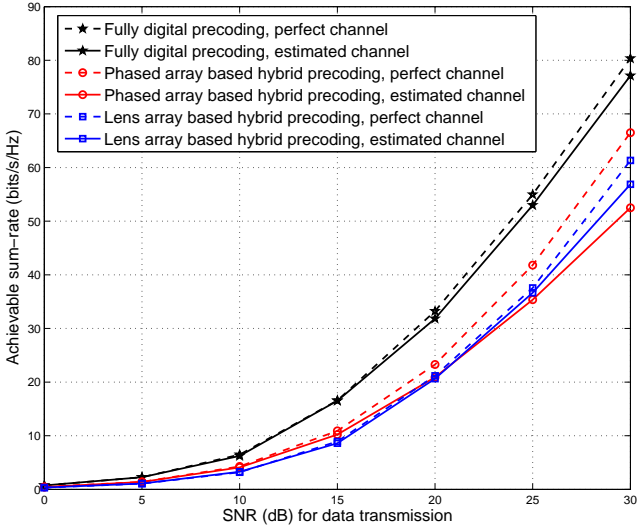}
\end{center}
\vspace*{-4mm}\caption{Comparison of sum-rate against the SNR for data transmission.} \label{FIG1}
\end{figure}

\begin{figure}[tp]
\begin{center}
\vspace*{-1mm}\includegraphics[width=0.95\linewidth]{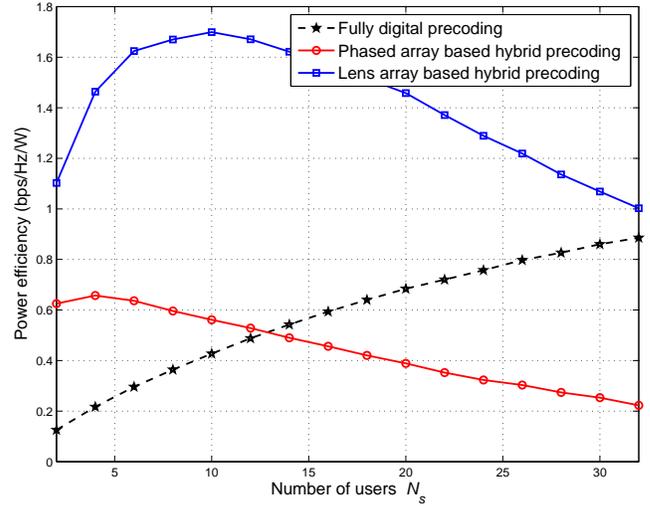}
\end{center}
\vspace*{-4mm}\caption{Comparison of power efficiency against the number of users ${{N_{\rm{s}}}}$.} \label{FIG1}
\end{figure}

\begin{figure}[tp]
\begin{center}
\vspace*{+1mm}\includegraphics[width=0.95\linewidth]{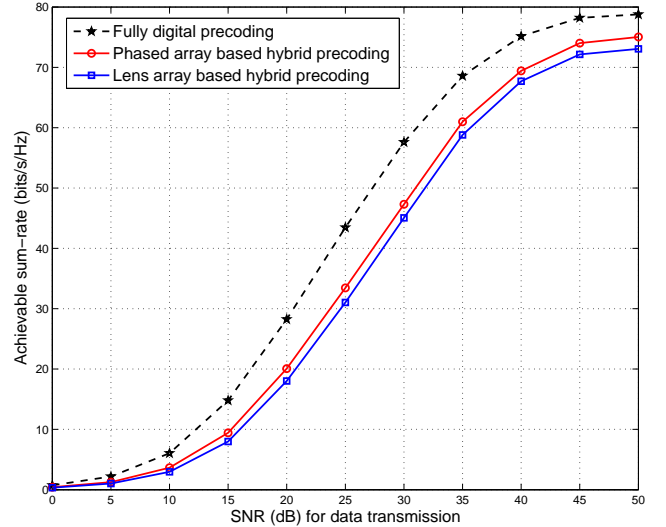}
\end{center}
\vspace*{-4mm}\caption{Comparison of sum-rate in multi-cell scenario.} \label{FIG1}
\end{figure}

Fig. 5 shows the comparison of power efficiency. We define the power efficiency ${\eta}$ as ${\eta  = R/\left( {{P_{\rm{T}}} + {P_{\rm{H}}}} \right)}$~\cite{amadorilow}, where ${R}$ is the achievable sum-rate, ${{{P_{\rm{T}}}}}$ is the transmission power, which can be set as ${{P_{\rm{T}}} = 2.5{\rm{W}}}$ (34 dBm) for outdoor mmWave MIMO in a small cell~\cite{rappaport2013millimeter}, ${{{P_{\rm{H}}}}}$ is the power consumed by hardware architecture. For  full-PAHP as shown in Fig. 2 (b), we have ${{P_{\rm{T}}} = N{P_{{\rm{A}}}} + {N_{{\rm{RF}}}}N{P_{{\rm{PS}}}} + {N_{{\rm{RF}}}}{P_{{\rm{SP}}}} + N{P_{{\rm{CO}}}} + {N_{{\rm{RF}}}}{P_{{\rm{RF}}}}}$, where ${{P_{{\rm{A}}}}}$, ${{P_{{\rm{PS}}}}}$, ${{P_{{\rm{SP}}}}}$, ${{P_{{\rm{CO}}}}}$, and ${{P_{{\rm{RF}}}}}$ are the power consumed by amplifier, phase shifter, power splitter, power combiner, and RF chain, respectively. For LAHP with the proposed adaptive selecting network as shown in Fig. 3 (b), we have ${{P_{\rm{T}}}\! =\! N{P_{{\rm{A}}}} + {N_{{\rm{RF}}}}N{P_{{\rm{SW}}}} + {N_{{\rm{RF}}}}{P_{{\rm{SP}}}} + N{P_{{\rm{CO}}}} + {N_{{\rm{RF}}}}{P_{{\rm{RF}}}}}$, where ${P_{{\rm{SW}}}}$ is the power consumed by switch. Finally, for fully digital precoding as shown in Fig. 2 (a), we have ${{P_{\rm{T}}} = N{P_{{\rm{A}}}} + N{P_{{\rm{RF}}}}}$. In this paper, some referenced values\footnote{Note that the power consumption of RF modules usually presents high variability, which depends on the specific implementation type and performance requirement~\cite{mendez2016hybrid}. In this paper, we just adopt the conservative (high) values, and the results in Fig. 5 can be considered as the lower bounds of the power efficiencies in practice.} are adopted as ${{P_{{\rm{A}}}}\! =\! 20{\rm{mW}}}$~\cite{mendez2016hybrid}, ${{P_{{\rm{SP}}}}\! =\! {P_{{\rm{CO}}}}\! =\! 10{\rm{mW}}}$~\cite{mendez2016hybrid}, ${{P_{{\rm{RF}}}} \!=\! 250{\rm{mW}}}$~\cite{heath2015overview}, ${{P_{{\rm{PS}}}} = 30{\rm{mW}}}$ for 4-bit phase shifter~\cite{mendez2016hybrid}, and ${{P_{{\rm{SW}}}} = 5{\rm{mW}}}$~\cite{mendez2016hybrid}. Fig. 5 shows that both PAHP and LAHP can achieve much higher power efficiency than fully digital precoding when the number of users is not large (e.g., ${{N_{\rm{s}}} \le 8}$). However, when ${{N_{\rm{s}}} > 12}$,  LAHP still enjoys high power efficiency, but PAHP performs even worse than fully digital precoding. This conclusion is contrary to the widely accepted misconception that PAHP is more power efficient than fully digital precoding, but consistent with a very recent analysis in~\cite{amadorilow,gao16bs,Hogan16}.  It can be explained by the fact that as ${{N_{\rm{s}}}}$ grows, the number of phase shifters required by PAHP increases rapidly. As a result, the power consumption of phase shifters will be huge, even higher than that of RF chains.

Next, we extend the performance comparison to multi-cell scenario as shown in Fig. 6, where the number of cells is set as 2 for simplicity but without loss of generality. Each cell has one BS with ${N = 256}$ antennas to serve ${K = 16}$ users with the same transmission power. The channels between the BS and the users in its own cell, and the channels between the BS and the users in the neighbor cell, are generated following the same model in Fig. 4. We assume that no cooperation between BSs exists, and each BS only knows the channels of the users in its own cell. From Fig. 6, we observe that the achievable sum-rates of all technologies will not grow without bound when the SNR for data transmission increases. This is due to the fact that high SNR also incurs high inter-cell interference. However, Fig. 6 still shows that both PAHP and LAHP can achieve the near-optimal performance, even in the multi-cell scenario with inter-cell interference.

%\section{Lessons and Future Research Directions}\label{S7}
%Next, we will highlight the lessons learnt from the designs of PAHP and LAHP, and introduce some future research directions.
%
%\subsection{Lessons}\label{S5.1}
%The recently proposed PAHP and LAHP are two promising low RF-complexity technologies for outdoor mmWave MIMO systems. Both of them can achieve the near-optimal performance with considerably reduced number of RF chains. However, PAHP and LAHP are realized by different hardware architectures, leading to different design roadmaps. PAHP employs phase shifters, and deals with the low-rank mmWave MIMO channel in the spatial domain. As a result, the design roadmap of PAHP is to relax the constant modulus constraint on phase shifter by considering the low-rank characteristic, and approximate the original non-convex optimization problem as a convex one.  By contrast, LAHP employs lens array and switches, and deals with the sparse mmWave MIMO channel in the angular domain (beamspace). Therefore, the design roadmap of LAHP is to employ a low-complexity search algorithm to select the most appropriate beams, and design the digital precoder based on the dimension-reduced beamspace channel.

\section{Conclusions}\label{S8}
In this  paper, we have introduced two promising low RF-complexity technologies for mmWave MIMO with large antenna array, i.e., PAHP and LAHP, in detail. We have also proposed an adaptive selecting network for LAHP with low hardware cost and power consumption, which can formulate the beamspace channel estimation as a sparse signal recovery problem, and considerably reduce the pilot overhead. Finally, we have provided the complete and systematic performance comparison between PAHP and LAHP. It shows that PAHP achieves higher achievable sum-rate than LAHP when the channel is perfectly known, but LAHP is more robust to the channel estimation error. It also shows that LAHP enjoys higher power efficiency than PAHP, since the phase shifter network is replaced by the low-cost lens array and switches.

Besides the discussions above, there are still some open issues on low RF-complexity technologies for outdoor mmWave MIMO systems. For example, most of the existing low RF-complexity technologies are designed for the narrowband and time-invariant channels. However, due to the large bandwidth and the high frequency, the mmWave MIMO channels are more likely to be broadband and time-varying, which incurs new challenges. Take the PAHP for example, ``broadband" means that the analog beamformer cannot be adaptively adjusted according to the frequency, leading to more difficulties in signal processing design, while ``time-varying" means that we need to re-estimate the channel and re-compute the hybrid precoder frequently, leading to high pilot overhead and computational complexity. Therefore, designing low RF-complexity technologies for broadband time-varying channels will be an urging problem to solve.

\bibliography{IEEEabrv,Gao1Ref}

\begin{IEEEbiography}
{Xinyu Gao} (S'14) received the B.E. degree of Communication Engineering from Harbin Institute of Technology, Heilongjiang, China in 2014. He is currently working towards Ph.D. degree in Electronic Engineering from Tsinghua University, Beijing, China. His research interests include massive MIMO and mmWave communications. He has received the national scholarship in 2015 and the IEEE WCSP Best Paper Award in 2016.
\end{IEEEbiography}

\begin{IEEEbiography}
{Linglong Dai} (M'11-SM'14) received the Ph.D. degree from Tsinghua University, Beijing, China, in 2011. He is currently an Associate Professor of Tsinghua University. His current research interests include massive MIMO, millimeter-wave communications, multiple access, and sparse signal processing. He has received five conference Best Paper Awards (IEEE ICC 2013, IEEE ICC 2014, WCSP 2016, IEEE ICC 2017, and IEEE VTC 2017 Fall).
\end{IEEEbiography}

\vspace*{-10mm}
\begin{IEEEbiography}
{Akbar M. Sayeed} (F'12) received the Ph.D. degree from the University of Illinois at Urbana-Champaign,  USA, in 1996. He is a Professor of Electrical and Computer Engineering at the University of Wisconsin-Madison. His research interests include wireless communications, statistical signal processing, communication and information theory, wireless channel modeling, time-frequency analysis,  and development of basic theory, system architectures, and prototypes for new wireless technologies.
\end{IEEEbiography}

\end{document}